\documentclass[resetfootnote, trackchanges]{aastex7}

\usepackage[T1]{fontenc} 
\usepackage[utf8]{inputenc} 
\usepackage{amsmath}
\usepackage{amssymb}
\usepackage{hyperref}

\received{October 8, 2026}
\accepted{November 11, 2026}

\begin{document}

\title{A Joint Study of Epoch Astrometry and Radial Velocities of Gaia-4b}

\newcommand{\vdag}{(v)^\dagger}
\newcommand\aastex{AAS\TeX}
\newcommand\latex{La\TeX}

\newcommand\tess{TESS}
\newcommand\kepler{\textit{Kepler}}
\newcommand\gaia{\textit{Gaia}}
\newcommand\kms{$\textrm{km/s}$}
\newcommand\ms{$\textrm{m/s}$}
\newcommand\cms{$\textrm{cm/s}$}

\newcommand\xx{\textbf{FINDME}}
\newcommand\teff{T$_{\rm{eff}}$}
\newcommand\vsini{$v$~sin~$i$}
\newcommand\logg{log \textit{g}}
\newcommand{\unit}[1]{\ensuremath{\, \mathrm{#1}}} 
\newcommand\earthmass{$M_{\oplus}$}
\newcommand\earthradius{$R_{\oplus}$}
\newcommand\earthinsol{$S_{\oplus}$}
\newcommand\solmass{$M_{\odot}$}
\newcommand\solradius{$R_{\odot}$}
\newcommand\specificflux{ergs $\rm{s^{-1} cm^{-2}} \AA^{-1}$}
\newcommand\flux{ergs $\rm{s^{-1} cm^{-2}}$}

\author[0000-0001-8401-4300]{Shubham Kanodia}
\affiliation{\Carnegie}
\email{skanodia@carnegiescience.edu}

\author[0000-0001-9596-7983]{Suvrath Mahadevan}
\affiliation{\PSUAA}
\affiliation{\PSUCEHW}
\email{suvrath@astro.psu.edu}

\author[0000-0001-7409-5688]{Gudmundur Stefansson}  
\affiliation{\Schmidt}
\email{g.k.stefansson@uva.nl}

\author[0009-0005-9862-2549]{Domenico Barbato}  
\affiliation{INAF – Osservatorio Astronomico di Padova, Vicolo dell’Osservatorio 5, I-35122, Padova, Italy}
\email{domenico.barbato@inaf.it}

\author[0000-0001-8720-5612]{Joe P.\ Ninan}  
\affil{Department of Astronomy and Astrophysics, Tata Institute of Fundamental Research, Homi Bhabha Road, Colaba, Mumbai 400005, India}
\email{joe.ninan@tifr.res.in}

\author[0000-0001-9662-3496]{William D. Cochran} 
\affil{McDonald Observatory and Department of Astronomy, The University of Texas at Austin}
\affil{Center for Planetary Systems Habitability, The University of Texas at Austin}
\email{wdc@astro.as.utexas.edu}

\author[0000-0002-3853-7327]{Rachel B. Fernandes} 
\affiliation{\PSUAA}
\affiliation{\PSUCEHW}
\email{rbf5378@psu.edu}

\author[0000-0003-4835-0619]{Caleb I. Ca\~nas} 
\affiliation{\Goddard}
\email{c.canas@nasa.gov}

\author[0000-0002-9082-6337]{Andrea S.J.\ Lin} 
\affiliation{\Caltech}
\email{asjlin@caltech.edu}

\author[0000-0003-1324-0495]{Leonardo A.\ Paredes} 
\affiliation{Steward Observatory, University of Arizona, 933 N.\ Cherry Ave, Tucson, AZ 85721, USA}
\email{lparedes@arizona.edu}

\author[0000-0002-4445-1845]{Matteo Pinamonti} 
\affiliation{INAF – Osservatorio Astrofisico di Torino, Via Osservatorio 20, I-10025, Pino Torinese, Italy}
\email{matteo.pinamonti@inaf.it}

\author[0000-0003-0149-9678]{Paul Robertson} 
\affiliation{Department of Physics \& Astronomy, The University of California, Irvine, Irvine, CA 92697, USA}
\email{paul.robertson@uci.edu}

\author[0000-0002-4046-987X]{Christian Schwab} 
\affil{School of Mathematical and Physical Sciences, Macquarie University, Balaclava Road, North Ryde, NSW 2109, Australia}
\email{mail.chris.schwab@gmail.com}

\author[0000-0002-7504-365X]{Alessandro Sozzetti} 
\affiliation{INAF – Osservatorio Astrofisico di Torino, Via Osservatorio 20, I-10025, Pino Torinese, Italy}
\email{alessandro.sozzetti@inaf.it}

\newcommand{\Goddard}{NASA Goddard Space Flight Center, 8800 Greenbelt Road, Greenbelt, MD 20771, USA}
\newcommand{\Carnegie}{Earth and Planets Laboratory, Carnegie Science, 5241 Broad Branch Road, NW, Washington, DC 20015, USA}
\newcommand{\Zurich}{Department of Astrophysics, University of Z\"urich, Winterthurerstrasse 190, 8057 Z\"urich, Switzerland}
\newcommand{\PSUAA}{Department of Astronomy \& Astrophysics, The Pennsylvania State University, 525 Davey Laboratory, University Park, PA 16802, USA}
\newcommand{\PSUCEHW}{Center for Exoplanets and Habitable Worlds, The Pennsylvania State University, 525 Davey Laboratory, University Park, PA 16802, USA}
\newcommand{\Caltech}{Department of Astronomy, California Institute of Technology, 1200 E California Blvd, Pasadena, CA 91125, USA}
\newcommand{\Schmidt}{Astrophysics \& Space Center, Schmidt Sciences, New York, NY 10011, USA}

\begin{abstract}
Gaia-4b is the first confirmed exoplanet discovered with Gaia astrometry, based on its identification as a possible exoplanet in the solution in  Gaia Data Release 3 and subsequent confirmation with radial velocity. In this work we present a \texttt{python} framework called \texttt{pyGaiafits} that can jointly analyze astrometric time series and radial velocities. We use this to combine the precise along-scan astrometric time series for Gaia-4 from the Gaia DR4 pre-release with 44 radial velocities (RVs) obtained over three years from four independent spectrographs (HPF, HARPS-N, NEID, and FIES). By jointly fitting these two complementary datasets that collectively span a 12-year baseline ($\sim$7 orbital periods of Gaia-4b), we refine the system parameters and determine the true mass of Gaia-4b to be $10.70\pm0.32$~$M_J$. Critically, we do not find significant evidence for the long-term quadratic RV trend previously reported in the literature, recovering $\ddot{\gamma} = 5.5^{+8.3}_{-6.7}\;\mathrm{m\,s^{-1}\,yr^{-2}}$ consistent with zero. We discuss how this discrepancy likely arises from a degeneracy between orbital period and long-term trends when RVs span only $\sim$1.5 orbital cycles --- a degeneracy that is broken by anchoring the ephemeris with early Gaia astrometry. 
\end{abstract}



\section{Introduction}

ESA's Gaia mission \citep{collaboration_gaia_2016} was launched in 2013, and with its successive data releases is providing precise astrometry for billions of stars. The most recent one --- Gaia Data Release 3 \citep[DR3;][]{gaia_collaboration_gaia_2023} --- included a catalogue of unresolved sources where the observed astrometric motion was inconsistent with a single star's 5 parameter fit (parallax, RA, Dec, proper motions in RA and Dec.). This non-single star (NSS) catalogue \citep{halbwachs_gaia_2023} was vetted to yield 72 promising candidates that could be substellar in nature \citep{holl_gaia_2023}.

Of the 72 candidates, 28 were KM dwarfs in the Northern hemisphere that were observed by \citet{stefansson_gaia-4b_2025} to spectroscopically rule out astrophysical false positives.  In the process, they also validated the substellar nature for Gaia-4b and Gaia-5b by combining RVs with the Gaia DR3 two-body solution. Gaia-4 is a 0.644 $\pm$ 0.025 \solmass, 4034 $\pm$ 77 K early M or late K dwarf, and hosts the first exoplanet discovered on the basis of a valid astrometric solution --- Gaia-4b. 

Since the epoch astrometry was not publicly available with DR3, for the joint astrometry + RV fit \citet{stefansson_gaia-4b_2025} used the covariance matrix from the tabulated NSS solution and adopted a joint likelihood function that combined the likelihood functions from RVs and astrometry. 

Subsequently it was also observed as part of a selected sample of 14 stars from the NSS catalogue from Gaia DR3 \citep{barbato_gaps_2026}. For Gaia-4, 17 RVs from HARPS-N \citep{cosentino_harps-n_2012} were combined with the published RVs from \citet{stefansson_gaia-4b_2025}. The RV analysis included a uniform period prior based on the nominal value from the NSS catalogue, and the minimum mass derived from the RVs was then converted to a true mass using the orbital inclination constraints from the NSS. For Gaia-4, they identified a Keplerian solution, and compared models with and without a long term quadratic trend. Based on the Bayesian Information criterion, they favour and adopt the model that includes a long term trend with a quadratic coefficient of $0.0006 \pm 0.0001$ m s$^{-1}$ d$^{-2}$.

In this work we combine the published RVs from \citet{stefansson_gaia-4b_2025} and \citet{barbato_gaps_2026}, three new RVs from HPF, as well as epoch astrometry from Gaia DR4 released as part of the DR4 pre-release\footnote{\url{https://www.cosmos.esa.int/web/gaia/dr4-prerelease}} in June 2026. One of the 12 sources in this pre-release was Gaia DR4 \texttt{source\_ID} 1457486023639239296 or Gaia-4. We jointly analyze all the RVs and epoch astrometry to place tight constraints on the Gaia-4b system parameters and do not find any evidence for a long term RV trend. In Section \ref{sec:observations} we describe the astrometric and RV observations used in this analysis followed by describing the model developed for analysis in Section \ref{sec:analysis}. We then discuss our results and conclude in Section \ref{sec:discussion}.

\begin{deluxetable*}{ccccc}
\tablecaption{Joint RV and astrometric orbit fit \label{tab:BinaryRVFit}}
\tablewidth{0pt}
\tabletypesize{\footnotesize}
\tablehead{
  \colhead{Parameter} &
  \colhead{Description} &
  \colhead{Value} &
  \colhead{Value} &
  \colhead{Value} \\
  \colhead{} &
  \colhead{} &
  \colhead{\citet{stefansson_gaia-4b_2025}\tablenotemark{b}} &
  \colhead{\citet{barbato_gaps_2026}\tablenotemark{c}} &
  \colhead{This work\tablenotemark{d}}
}
\startdata
\multicolumn{5}{l}{\hspace{-0.2cm} \textbf{Orbital Parameters:}}           \\
$P$\dotfill      & Orbital period (days) \dotfill               & $571.3^{+1.4}_{-1.3}$    & $588.58^{+2.15}_{-2.50}$ & $575.9\pm0.9$ \\
                 & Orbital period (yr) \dotfill                 & \nodata                  & \nodata                  & $1.577\pm0.002$ \\
$t_{\rm peri}$\dotfill & Time of periastron (days)\tablenotemark{a} \dotfill & $-550.775^{+8.5}_{-8.4}$ & \nodata & $37.7\pm6.4$ \\
$e$\dotfill      & Eccentricity \dotfill                        & $0.338^{+0.026}_{-0.023}$ & $0.318^{+0.011}_{-0.012}$ & $0.319\pm0.012$ \\
$\omega$\dotfill & Argument of periastron ($^{\circ}$) \dotfill & $180.3^{+5.5}_{-4.8}$    & $-159.52\pm2.84$         & $202.4\pm3.7$ \\
$\Omega$\dotfill & Longitude of ascending node ($^{\circ}$) \dotfill & $158.6\pm5.1$       & \nodata                  & $172.6\pm2.1$ \\
$i$\dotfill      & Inclination ($^{\circ}$) \dotfill            & $116.9\pm4.4$            & $115.44\pm4.60$ & $120.5\pm1.6$ \\
$K$\dotfill      & RV semi-amplitude (m\,s$^{-1}$) \dotfill     & $368.0^{+22.0}_{-18.0}$  & $347.42^{+2.15}_{-2.75}$ & $315.1\pm3.8$ \\
\multicolumn{5}{l}{\hspace{-0.2cm} \textbf{Astrometric parameters:}} \\
$a_0$\dotfill    & Astrometric amplitude (mas) \dotfill         & $0.279^{+0.016}_{-0.014}$ & $0.31\pm0.04$ & $0.250\pm0.004$ \\
$\varpi$\dotfill & Parallax (mas) \dotfill                      & \nodata                  & \nodata                  & $13.6251\pm0.0074$ \\
$\mu_{\alpha*}$\dotfill & Proper motion in R.A. (mas\,yr$^{-1}$) \dotfill  & \nodata    & \nodata                  & $-75.533\pm0.004$ \\
$\mu_{\delta}$\dotfill  & Proper motion in Decl. (mas\,yr$^{-1}$) \dotfill & \nodata    & \nodata                  & $17.966\pm0.004$ \\
$\log \sigma_{\rm astro}$\dotfill & Natural log astrometric jitter (mas) \dotfill & \nodata     & \nodata                  & $-3.00^{+0.08}_{-0.09}$ \\
\multicolumn{5}{l}{\hspace{-0.2cm} \textbf{Derived:}}           \\
$A$\dotfill      & Thiele-Innes coefficient (mas) \dotfill      & $0.258\pm0.018$          & \nodata                  & $0.2228^{+0.0079}_{-0.0088}$ \\
$B$\dotfill      & Thiele-Innes coefficient (mas) \dotfill      & $-0.103\pm0.024$         & \nodata                  & $-0.078\pm0.010$ \\
$F$\dotfill      & Thiele-Innes coefficient (mas) \dotfill      & $-0.047\pm0.026$         & \nodata                  & $-0.110\pm0.014$ \\
$G$\dotfill      & Thiele-Innes coefficient (mas) \dotfill      & $-0.116\pm0.023$         & \nodata                  & $-0.104\pm0.009$ \\
$m_2 \sin i$\dotfill & Minimum companion mass ($M_{\rm J}$) \dotfill & \nodata           & $10.30\pm0.45$           & $9.21\pm0.26$ \\
$m_2$\dotfill    & Companion mass ($M_{\rm J}$) \dotfill        & $11.8^{+0.73}_{-0.66}$   & $11.41\pm0.69$           & $10.70\pm0.32$ \\
\multicolumn{5}{l}{\hspace{-0.2cm} \textbf{RV Trend:}}           \\
$\dot{\gamma}$\dotfill  & Linear RV trend (m\,s$^{-1}$\,yr$^{-1}$) \dotfill    & \nodata & $-76.7\pm21.9$ &  $-9.0\pm7.1$ \\
$\ddot{\gamma}$\dotfill & Quadratic RV trend (m\,s$^{-1}$\,yr$^{-2}$) \dotfill & \nodata & $80.0\pm13.3$  & $5.5^{+8.3}_{-6.7}$ \\
\multicolumn{5}{l}{\hspace{-0.2cm} \textbf{RV Parameters:}}           \\
$\gamma_{\rm HPF}$\dotfill     & RV offset, HPF (m\,s$^{-1}$) \dotfill     & $-55.6^{+10.0}_{-9.2}$   & $-53.67^{+7.84}_{-7.57}$   & $-43.1^{+9.2}_{-11.1}$ \\
$\gamma_{\rm HARPS-N}$\dotfill & RV offset, HARPS-N (m\,s$^{-1}$) \dotfill & \nodata                  & $212.90^{+8.04}_{-7.33}$   & $222.8\pm5.6$ \\
$\gamma_{\rm NEID}$\dotfill    & RV offset, NEID (m\,s$^{-1}$) \dotfill    & $-48.0\pm13.0$           & $-33.52^{+11.07}_{-10.61}$ & $-23.7\pm8.1$ \\
$\gamma_{\rm FIES}$\dotfill    & RV offset, FIES (m\,s$^{-1}$) \dotfill    & $67.1\pm10.0$            & $61.13^{+10.05}_{-9.94}$   & $70.8\pm15.0$ \\
$\sigma_{\rm HPF}$\dotfill     & RV jitter, HPF (m\,s$^{-1}$) \dotfill     & $17.0\pm11.0$            & $15.12^{+10.14}_{-7.65}$   & $17.6^{+10.0}_{-7.9}$ \\
$\sigma_{\rm HARPS-N}$\dotfill  & RV jitter, HARPS-N (m\,s$^{-1}$) \dotfill & \nodata                  & $8.29^{+2.83}_{-1.83}$     & $12.2^{+3.4}_{-2.5}$ \\
$\sigma_{\rm NEID}$\dotfill    & RV jitter, NEID (m\,s$^{-1}$) \dotfill    & $23.5^{+10.0}_{-6.6}$    & $19.57^{+9.49}_{-6.16}$    & $16.7^{+8.7}_{-6.2}$ \\
$\sigma_{\rm FIES}$\dotfill    & RV jitter, FIES (m\,s$^{-1}$) \dotfill    & $2.5^{+13.0}_{-2.3}$     & $10.40^{+14.83}_{-7.25}$   & $23.0^{+16.5}_{-12.1}$ \\
\enddata
\tablenotetext{a}{Reported as an offset from $t_{\rm ref}$ for Gaia DR4 = 2017.5 = 2457936.875.}
\tablenotetext{b}{Gaia\,+\,RV (adopted) solution based on the Thiele-Innes coefficients from \citet{holl_gaia_2023}}
\tablenotetext{c}{Values for PM~J13580+3141; the ``$_{\rm b}$'' (joint RV\,+\,astrometry) solution is quoted in preference to the Gaia NSS solution where both are given.}
\tablenotetext{d}{From a combined fit constrained by both RVs + astrometry and described in Section \ref{sec:mass}.}
\tablecomments{Uncertainties are the 16th--84th percentiles of the posterior. Inclination $>90^{\circ}$ indicates retrograde motion on the sky.}
\end{deluxetable*}
\section{Observations}\label{sec:observations}

\subsection{Gaia DR4 pre-release}\label{sec:gaiaobs}
\begin{figure*}[!t] 
\centering
\includegraphics[width=\textwidth]{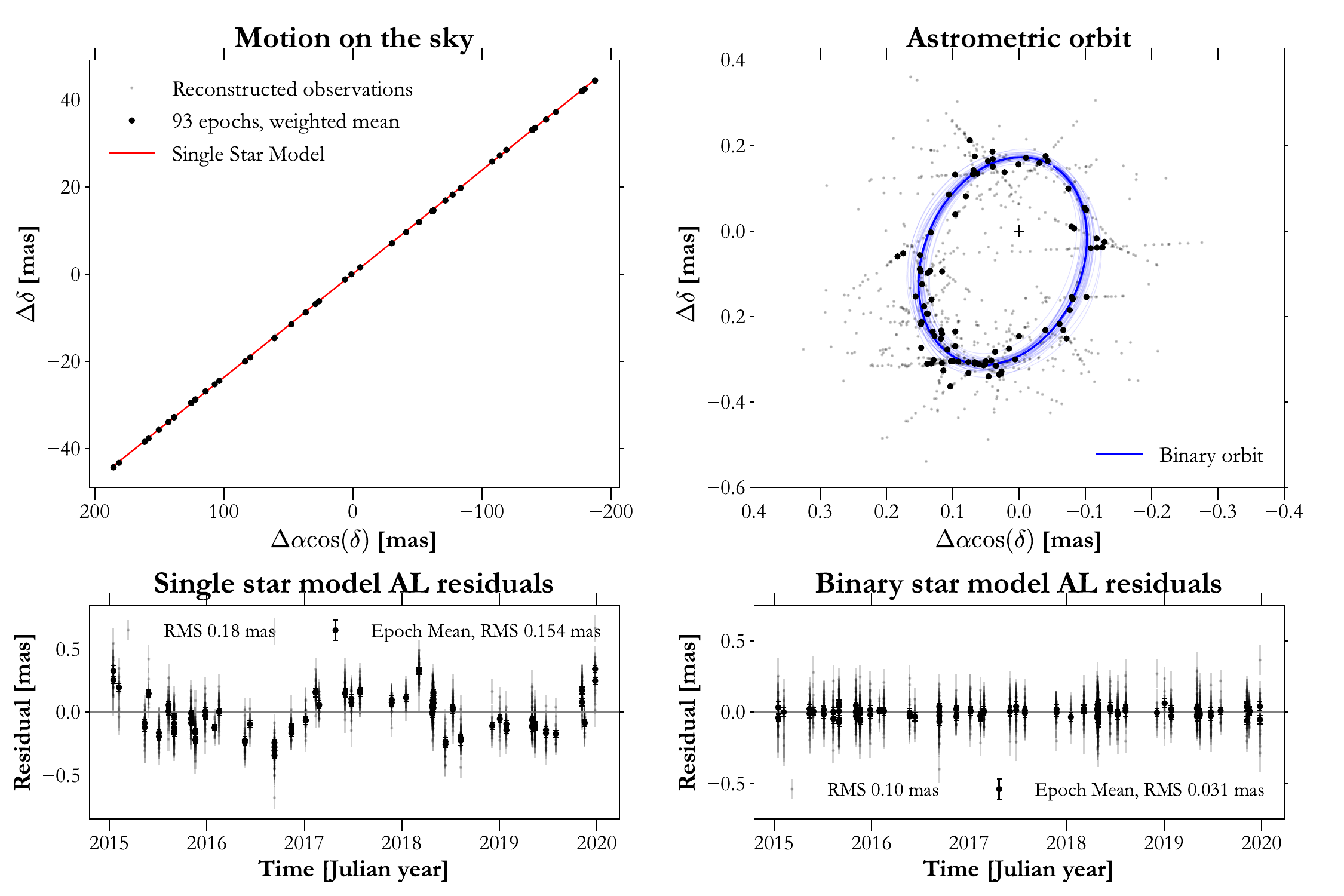}
\caption{\small  Gaia DR4 pre-release astrometric time series. \textbf{Top Left:} The reconstructed on-sky motion for Gaia-4 based on the best-fit proper motion. Note that the effects of parallax are not shown here. The 93 observations binned by epoch are shown as solid black dots, whereas the individual CCD observations are shown as small grey dots. \textbf{Bottom Left:} Residuals from the single star fit for the along scan measurements. \textbf{Top Right:} The astrometric orbit of the photocenter after subtracting the parallax and proper motion. As before, the individual CCD observations are shown as small grey dots, and binned observations as solid black dots. The $+$ marks the center of mass, while the solid blue line is the orbit derived from the median value of the orbital parameters. We also plot 40 random draws from our posterior sampling for illustrative purposes. \textbf{Bottom Right:} Residuals for the binary star fit for the along scan measurements.}\label{fig:Astrometry}
\end{figure*}

The epoch astrometry\footnote{A draft description of the DR4 data product is included here \url{https://www.cosmos.esa.int/web/gaia/dr4\#datamodel} .} consisted of 1150 observations across 115 field of view (FOV) transits for each of the 10 CCDs (1 Sky Mapper CCD + 9 Astrometric Field), of which 1077 contain finite along-scan (AL) abscissa measurements ($w$; \texttt{centroid\_pos\_al}), errors  ($\sigma_w$; \texttt{centroid\_pos\_error\_al}), scan-position angle $\theta$, and AL parallax factor ($f_{\rm AL}$). Of these we limit our analysis to the 824 observations marked True for \texttt{used\_by\_agis\_al} in DR4 \citep{lindegren_astrometric_2012} that are taken across 93 unique transits and shown in \autoref{fig:Astrometry}. We retain the individual CCD astrometric measurements in the likelihood and bin only for display. The AL abscissa is a one-dimensional projection of the sky position onto the instantaneous scan direction and the 824 observations have a median error of 0.082 mas. All model fits are thus performed in AL space rather than on reconstructed two-dimensional positions.

\subsection{Spectroscopic Observations}
We use the 24 literature radial velocity (RV) observations from \citet{stefansson_gaia-4b_2025}, which include 11 visits from from the Habitable-zone Planet Finder \citep{mahadevan_habitable-zone_2012, mahadevan_habitable-zone_2014} on the 10.0 m Hobby Eberly Telescope\footnote{Based on observations obtained with the Hobby-Eberly Telescope (HET), which is a joint project of the University of Texas at Austin, the Pennsylvania State University, Ludwig-Maximillians-Universitaet Muenchen, and Georg-August Universitaet Goettingen. The HET is named in honor of its principal benefactors, William P. Hobby and Robert E. Eberly.}  \citep{ramsey_early_1998, hill_hetdex_2021}, 8 from NEID \citep{halverson_comprehensive_2016, schwab_design_2016} on the 3.5 m WIYN telescope\footnote{The WIYN Observatory is a joint facility of the NSF’s National Optical-Infrared Astronomy Research Laboratory, Indiana University, the University of Wisconsin-Madison, Pennsylvania State University, Purdue University and Princeton University.} and 5 from the FIber-fed Echelle Spectrograph \citep[FIES;][]{telting_fies_2014} mounted to the 2.56 m Nordic Optical Telescope. We then include 17 RVs from High Accuracy Radial velocity Planet Searcher in the Northern hemisphere \citep[HARPS-N;][]{cosentino_harps-n_2012} on the 3.6 m TNG telescope.

We also add three more visits from HPF from 2025 April 9, 2026 July 24, and 2026 July 29. Like previous HPF exposures these were reduced using the HPF pipeline following the procedure described in \citep{ninan_habitable-zone_2018, kaplan_monitoring_2019, metcalf_stellar_2019}. We then performed barycentric corrections using \texttt{barycorrpy} \citep{kanodia_python_2018} and subsequently extracted precise RVs from the 1D spectra using the template matching algorithm \citep[SERVAL;][]{zechmeister_spectrum_2018} modified for HPF \citep{stefansson_neptune-mass_2023}. SERVAL builds an empirical template using all the observations of the star that is used to derive shifts (velocities) from individual epoch spectra. Since we are using a larger set of HPF observations here than those presented in \citet{stefansson_gaia-4b_2025} and re-deriving the empirical template, our HPF RVs also include minor changes ($\ll$ median RV errorbars) to the HPF RVs associated with the epochs presented therein. These new RVs extend the baseline on previously published HPF RVs without any instrument breaks in the interim, and hence do not necessitate an RV offset in the model.

\begin{figure*}[!t] 
\centering
\includegraphics[width=\textwidth]{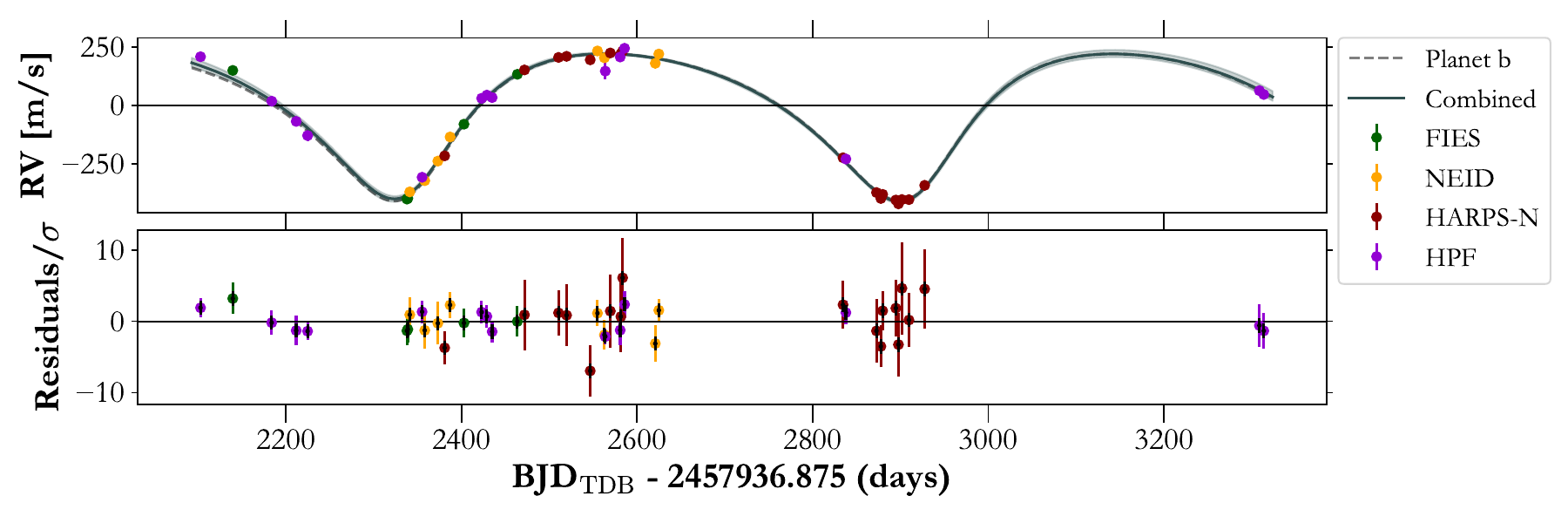}
\caption{\small  Time series RVs of Gaia-4 taken with FIES, NEID, HARPS-N, HPF. The best-fitting model derived from the joint fit to the DR4 pre-release astrometric time series and RVs is also shown, where the dashed line is the Keplerian for planet b and the solid line shows the model including the RV trend.  The model includes the 16-84$\%$ confidence interval in a lighter shade.  The bottom panel shows the residuals after subtracting the model from the data. The RV time series shown here is included as data behind the figure with this manuscript.  Instrument errorbars are shown in black, while RV jitter added in quadrature is shown in the background on each point in colour. We reference the time series against Gaia DR4 epoch  J2017.5. }\label{fig:RVTimeSeries}
\end{figure*}

\begin{figure}[!t] 
\centering
\includegraphics[width=0.75\textwidth]{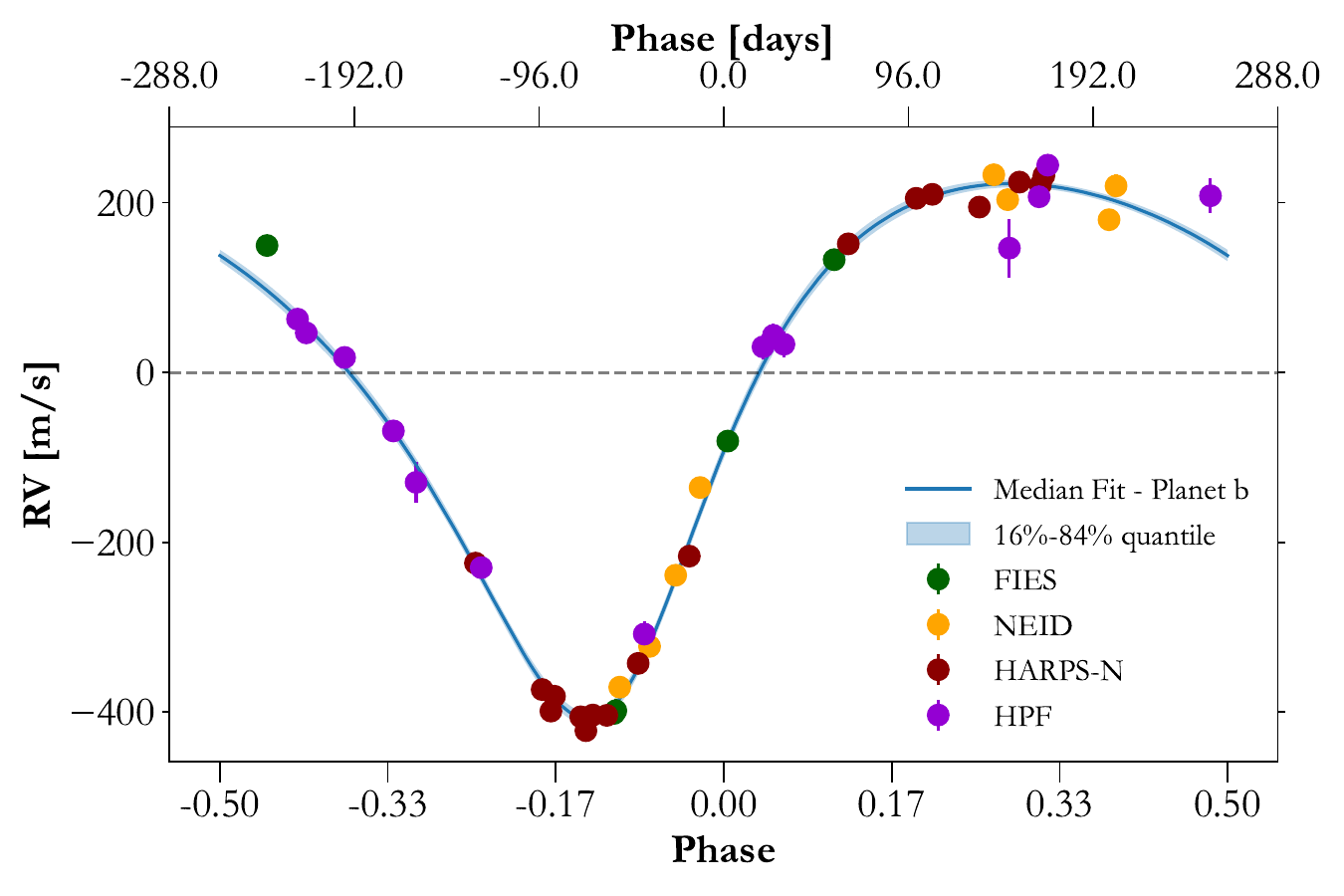}
\caption{\small RV observations phase-folded to the best fit orbital period from the joint fit described in Section \ref{sec:analysis}. Instrument errorbars are shown in colour.}\label{fig:RVPhaseCurve}
\end{figure}

\section{Analysis}\label{sec:analysis}

\subsection{Single-star astrometric model}\label{sec:ssmodel}

The single-star model comprises the standard five astrometric parameters: position offsets $\Delta\alpha^{*}$ and $\Delta\delta$ relative to the catalogue reference position, proper motions, $\mu_{\alpha^{*}}$ and $\mu_{\delta}$, and parallax $\varpi$ \citep{esa_hipparcos_1997, lindegren_gaia_2020}. These are projected onto the scan direction to predict the AL abscissa at epoch $t$ using the scan position angle $\theta$ as described in Section \ref{sec:gaiaobs},
\begin{equation}
w_{\rm 5p}(t) = \left(\Delta\alpha^{*} + \mu_{\alpha^{*}}\,\Delta t\right)\sin\theta
              + \left(\Delta\delta + \mu_{\delta}\,\Delta t\right)\cos\theta
              + \varpi\,f_{\rm AL}(t),
\label{eq:5p}
\end{equation}
where $\Delta t = t - t_{\rm ref}$ is measured from the DR4 reference epoch (2017.5 = JD 2457936.875) and $f_{\rm AL}$ is the AL parallax factor supplied with the epoch data.

\subsection{Astrometric orbit}\label{sec:orbitmodel}

For the binary model we add the photocenter orbit to Equation~\ref{eq:5p}. The Keplerian orbit is described by the Campbell elements: period $P$, photocentric semi-major axis $a_0$, eccentricity $e$, argument of periastron $\omega$, longitude of the ascending node $\Omega$, inclination $i$, and time of periastron passage $t_{\rm peri}$. We evaluate  the orbit with \texttt{exoplanet} \citep{foreman-mackey_exoplanet_2021}, which solves Kepler's equation and returns the relative position in the orbital plane projected onto the sky. This projected orbit is likewise projected on to the along-scan direction,
\begin{equation}
w_{\rm orb}(t) = \Delta\alpha^{*}_{\rm orb}(t)\,\sin\theta
               + \Delta\delta_{\rm orb}(t)\,\cos\theta,
\end{equation}
with $\Delta\delta$ measured toward North and $\Delta\alpha^{*}$ toward
East, and the full binary model is $w_{\rm bin} = w_{\rm 5p} + w_{\rm orb}$. The Thiele-Innes constants are retained as derived quantities for comparison with the DR3 NSS catalogue solutions.

\begin{align}
A &= a_0\,(\cos\omega\cos\Omega - \sin\omega\sin\Omega\cos i), \nonumber \\
B &= a_0\,(\cos\omega\sin\Omega + \sin\omega\cos\Omega\cos i), \nonumber \\
F &= -a_0\,(\sin\omega\cos\Omega + \cos\omega\sin\Omega\cos i), \nonumber \\
G &= -a_0\,(\sin\omega\sin\Omega - \cos\omega\cos\Omega\cos i),
\label{eq:TI}
\end{align}
which satisfy $A^2 + B^2 + F^2 + G^2 = a_0^2\,(1 + \cos^2 i)$. We note that the Thiele-Innes coefficients are invariant under the simultaneous transformation $(\omega,\Omega)\rightarrow(\omega+\pi,\,
\Omega+\pi)$, and hence an astrometry-only solution cannot solve this degeneracy, but an astrometry + RV model can.

Given the large potential prior space for sampling the Campbell elements directly, we sample in $\ln P$, $\ln a_0$, and $\cos i$. Furthermore, we sample $\Omega$ and the orbital phase as angles using the  \texttt{pmx.Angle} reparametrisation\footnote{The function \texttt{pmx.Angle()} can be found \href{https://github.com/exoplanet-dev/pymc-ext/blob/c2696b3c9244661a3e8e240c22d2f1193c561922/src/pymc3_ext/distributions/transforms.py\#L87}{here} and is described in the \texttt{pymc3-ext} \href{https://github.com/exoplanet-dev/pymc-ext/blob/c2696b3c9244661a3e8e240c22d2f1193c561922/README.md/\#distributions}{documentation}.}. within \texttt{pymc3-ext} \citep{foreman-mackey_exoplanet_2021}, which imposes a uniform prior while avoiding the discontinuity at $\pm\pi$ by sampling from a two-dimensional space proportional to \textit{sine} and \textit{cosine}. To improve the sampling for eccentric orbits, we sample eccentricity and argument of periastron jointly as the unit-disk vector $(\sqrt{e}\cos\omega,\,\sqrt{e}\sin\omega)$ using \texttt{pmx.UnitDisk}\footnote{The function \texttt{pmx.UnitDisk()} can be found \href{https://github.com/exoplanet-dev/pymc-ext/blob/c2696b3c9244661a3e8e240c22d2f1193c561922/src/pymc3_ext/distributions/transforms.py\#L61}{here} and is described in the \texttt{pymc3-ext} \href{https://github.com/exoplanet-dev/pymc-ext/blob/c2696b3c9244661a3e8e240c22d2f1193c561922/README.md/\#distributions}{documentation}.} within \texttt{pymc3-ext} \citep{foreman-mackey_exoplanet_2021}.


\subsection{Joint radial velocity model}\label{sec:rvmodel}

The RVs are modelled with the same Keplerian orbit, sharing $P$, $t_{\rm peri}$, $e$, and $\omega$ with the astrometric solution. We perform two fits, which are described further in Section \ref{sec:mass}: in the first, the astrometric amplitude $a_0$ is sampled independently from the RV semi-amplitude $K$; in the second, $K$ is additionally constrained by $a_0$.


Each RV instrument $j$ carries its own systemic offset $\gamma_j$ and jitter $\sigma_{{\rm jit},j}$, added in quadrature to the reported uncertainties. A systemic acceleration common to all RV instruments is included as a quadratic polynomial in time relative to the mean RV epoch.  The astrometric model also constrains the inclination $i$ and $\Omega$, and includes a log astrometric jitter term. 

\subsection{Likelihood and Inference}\label{sec:likelihood}

Both datasets are modelled with independent Gaussian likelihoods. For the astrometry an AL jitter term $\sigma_{\rm AL}$ is added in quadrature to the formal per-observation uncertainties to absorb unmodelled epoch-to-epoch systematics, and analogously for the RVs with the per-instrument jitter. The total log-likelihood is the sum of the two individual log-likelihoods.

The model is implemented in \texttt{PyMC3} \citep{salvatier_probabilistic_2016} using \texttt{exoplanet} \citep{foreman-mackey_exoplanet_2021} and \texttt{Theano} \citep{the_theano_development_team_theano_2016}. We first fit the single-star model alone, and use its maximum a posteriori (MAP) solution to initialise the five astrometric parameters of the binary fit. While this was not an issue for Gaia-4 due to its large astrometric amplitude and astrometric sampling over almost four orbital periods, without this staging the optimiser can settle in a local mode for longer period orbits in which the orbital amplitude collapses toward zero and the astrometric signal is absorbed entirely by the proper motion. The MAP optimisation itself proceeds in stages, first varying the five astrometric parameters and the jitter, then releasing all parameters jointly.

We run the sampling with 3 chains of 3000 tuning steps and 1500 draws. Convergence is assessed with Gelman-Rubin statistic \citep[$\hat{\text{R}} \le 1.01$;][]{ford_improving_2006}. We release this framework in a publicly accessible GitHub repository called \texttt{pyGaiafits}\footnote{\url{https://github.com/shbhuk/pyGaiafits} with the version used here stored on \url{https://zenodo.org/records/23922472}.}. Unlike affine-invariant samplers like \texttt{emcee} \citep{foreman-mackey_emcee_2013}, \texttt{pyGaiafits} samples the posterior using the Hamiltonian Monte Carlo method \citep[HMC;][]{duane_hybrid_1987, neal_mcmc_2012}, specifically the No-U-Turn Sampler \citep[NUTS;][]{hoffman_no-u-turn_2014}, implemented in \texttt{PyMC3} alongside the Keplerian solver used in \texttt{exoplanet}. HMC scales more favourably with the dimensionality of a joint astrometry + RV fit, where the number of parameters grow quickly with multiple instruments. The framework is also trivial to expand to include additional cases that already exist within the \texttt{exoplanet} package including Gaussian processes \citep{rasmussen_gaussian_2006}, photometric rotation, etc. Furthermore, our implementation is deliberately minimal, where the entire model (aside from utility functions) is contained in a single script and revealed to the user so that the priors and posteriors are legible and easy to inspect (unlike a black-box).

\subsection{Companion mass}\label{sec:mass}
 
The RVs measure the semi-amplitude ($K$) that constrain the mass function of the system,
\begin{equation}
f(M) \equiv \frac{M_2^3\sin^3 i}{\left(M_1+M_2\right)^2}
      = \frac{P\,K^3\left(1-e^2\right)^{3/2}}{2\pi G},
\label{eq:specmassfunc}
\end{equation}
which for a single-lined spectroscopic binary is degenerate with the inclination. The astrometry breaks that degeneracy by measuring $i$ directly, so that Equation~\ref{eq:specmassfunc} may be solved for the companion mass ($M_2$) given an independent estimate of the primary mass $M_1$. We adopt $M_1 = 0.644 \pm 0.025$ \solmass{} from \citet{stefansson_gaia-4b_2025}, and solve the resulting cubic iteratively. Uncertainties are propagated by drawing $\left(P, K, e, i\right)$ jointly from the posterior chains, and drawing $M_1$ from an independent Gaussian; we quote the median and the 16th--84th percentiles of the resulting $M_2$ distribution. 

We can also estimate an astrometric mass function for the object that is separate from the above spectroscopic mass function. This is deliberately done since the lack of consistency between the two would suggest a non-dark, i.e., luminous secondary component. The relative separation ($a$) is $a_1 + a_2$, where $a_1, a_2$ are the semi-major axes of the primary and secondary; additionally  we know that $a_1 M_1 = a_2 M_2$ since the two bodies orbit the common center of mass. So

\begin{align}
    a &= a_1\left(1 + \frac{M_1}{M_2} \right) \nonumber \\
    \intertext{which reduces Kepler's third law (with \textit{a} in AU, \textit{P} in years, and \textit{M} in \solmass) to}
    a^3 &= (M_1 + M_2) P^2 \Rightarrow a_1 = M_2 (M_1 + M_2)^{-2/3} P^{2/3}, \nonumber \\
    \intertext{then a similar astrometric mass function can be derived,}
    \frac{a_1^3}{P^{2}} &= \frac{M_2^3}{(M_1 + M_2)^{2}}   \label{eq:astromassfunction}
\end{align}

Assuming $a_1$ is the astrometric amplitude ($a_0$) we can solve \autoref{eq:astromassfunction} iteratively, similar to \autoref{eq:specmassfunc} and obtain $M_2$. In the first of these fits, where $a_0$ and $K$ are sampled independently, the spectroscopic amplitude from \autoref{eq:specmassfunc} gives a companion mass of 10.82~$\pm$~0.45~$M_J$, while the astrometric amplitude gives a mass of 10.57~$\pm$~0.43~$M_J$, which are consistent. 

Similarly, instead of using the astrometric amplitude, we can also obtain $a_1$ from the spectroscopic orbit alone; we denote this determination ($a_{0,b}$),
\begin{equation}
a_{1}\sin i = a_{0,b}\sin i = \frac{K\,P\,\sqrt{1-e^2}}{2\pi},
\label{eq:a1sini}
\end{equation}
which, combined with the astrometric inclination and parallax, may be expressed in angular units and compared directly with the measured photocentric semi-major axis. In general the two differ,
\begin{equation}
a_0 = a_{0,b}\left(1 - \frac{\beta}{B}\right)
    \simeq a_{0,b}\left(1-\frac{\beta M_1}{M_2}\right) \quad \text{for } M_2 \ll M_1,
\label{eq:photocenter}
\end{equation}
where $B = M_2/(M_1+M_2)$ is the mass fraction and $\beta = L_2/(L_1+L_2)$ is the fractional flux contributed by the companion in the \gaia\ $G$ band. We deliberately include this redundancy while parametrizing the astrometric and RV orbits, where agreement between $a_0$ and $a_{0,b}$ indicates $\beta \simeq 0$, i.e.\ that the companion is dark and the photocentre traces the barycentric motion of the primary alone. Conversely, a  significant deficit in $a_0$ would instead imply a luminous companion partially dragging the photocentre toward the barycentre. For Gaia-4, we get  $a_0 = 0.247 \pm 0.008$ mas against $a_{0,b} = 0.253 \pm 0.008$ mas from \autoref{eq:a1sini}, a
difference of $0.005 \pm 0.013$ mas ($0.4\sigma$), corresponding to $\beta/B = 0.02^{+0.05}_{-0.05}$ that confirms that Gaia-4b is indeed dark within our measurement precision and that the photocenter corresponds to the reflex motion of the primary alone.

Having established this, we then repeat the fit imposing $a_1 = a_0$ as an additional  constraint on $K$ in \autoref{eq:a1sini}, instead of leaving the two independent. This tightens the inclination constrain from from $121.4 \pm 2.6^{\circ}$ to $120.5 \pm 1.6^{\circ}$, the photocentric amplitude from $0.247 \pm 0.008$ to $0.250 \pm 0.004$ mas, giving a final companion mass of $10.70\pm0.32$~$M_J$. We adopt this second fit for the parameters reported in \autoref{tab:BinaryRVFit}. Both modes are available in \texttt{pyGaiafits} through the \texttt{DeriveKfromA0} flag and described in the documentation.

 We show the models derived for astrometric time series in \autoref{fig:Astrometry}, while the RV time series and phase curves are shown in \autoref{fig:RVTimeSeries} and \autoref{fig:RVPhaseCurve} respectively. Importantly we do not recover a significant long term linear or quadratic RV trend (mentioned in \autoref{tab:BinaryRVFit}). We also validate our methodology and pipeline by running the same dataset using the \texttt{kima} python package \citep{faria_kima_2018} that uses the Diffusive Nested Sampling algorithm \citep{brewer_diffusive_2011} and has recently been updated to also allow joint analysis of RVs and Gaia astrometry. We obtain comparable posteriors with this fit ($< 1 \sigma$). We discuss and compare our results with previous studies in the next section.

\section{Discussion \& Conclusion}\label{sec:discussion}

Gaia DR4 astrometry is expected to contribute to the detection of 1000s of exoplanets, including 100s - 1000s of giant exoplanets around M-dwarfs \citep{perryman_astrometric_2014, sozzetti_astrometric_2014, lammers_exoplanet_2025}. A significant source of astrophysical false positives in detected planet candidates is expected to be unresolved binary stars (typically near-equal mass binaries), which would require spectroscopic follow-up to rule out \citep{marcussen_spectroscopic_2023, stefansson_gaia-4b_2025, marcussen_radial_2026}. Furthermore, the longest period orbits that are incomplete in DR4 will also necessitate RV follow-up, to rule out false positives and both provide an independent measurement and refine the orbital parameters. Gaia DR4 will include RVs from the Gaia Radial Velocity Spectrometer \citep[RVS;][]{cropper_gaia_2018} --- albeit not at the same precision as those used here --- that will help anchor and extend the RV time series that can be obtained for candidates for massive brown dwarfs post their detection from DR4 astrometry ($>$ December 2026).

In this work with Gaia-4b we illustrate this complementarity between astrometry and RVs. The DR4 astrometry spans 5.5 years between 25 July 2014 (10:30 UTC) to 20 January 2020 (22:00 UTC)\footnote{\url{https://www.cosmos.esa.int/web/gaia/dr4}}, whereas our RVs go from April 2023 to July 2026. The two datasets do not overlap and their midpoints are separated by $>$ 7.5 years or almost 5 orbital periods. The combined baseline is 12 years or $\sim$ 7 orbital periods, and it is this lever arm that drives the improvement in its ephemeris.  

This analysis, to our knowledge is only the second published work, after the discovery of Gaia-BH3 \citep{panuzzo_discovery_2024}, to jointly fit Gaia astrometric time series with radial velocities.  We also publicly release our analysis framework called \texttt{pyGaiafits}. Previous treatments of Gaia-4 necessarily worked with derived data products: \citet{stefansson_gaia-4b_2025} sampled the tabulated DR3 two-body solution through its covariance matrix, since the epoch astrometry was not public with DR3, while \citet{barbato_gaps_2026} used the NSS catalogue period as a prior, along with orbital inclination to obtain the true mass from the minimum mass.

\autoref{tab:BinaryRVFit} summarizes our system parameters alongside the previous studies. Not only do we refine the precision in orbital parameters, we also find the mass of Gaia-4b to be $10.70\pm0.32$$M_J$. Furthermore, the agreement between the spectroscopy and astrometry derived masses for Gaia-4b (Section \ref{sec:mass}) suggests that it is a `dark' companion that does not contribute luminosity to the system. Together, these reconfirm its planetary nature as the first exoplanet discovered and confirmed on the basis of a valid astrometric solution. We also touch upon some salient features of our results below.


\subsection{Absence of a long-term radial velocity trend}\label{sec:trend}
\citet{barbato_gaps_2026} analyse the RVs for Gaia-4b and compare the evidence from Keplerian models with and without a long-term quadratic RV component, to favour the model that includes one with a quadratic coefficient of $80.0 \pm 13.3\;\mathrm{m\,s^{-1}\,yr^{-2}}$ ($6\sigma$). In our joint fit we recover $\ddot{\gamma} = 5.5^{+8.3}_{-6.7}\;\mathrm{m\,s^{-1}\,yr^{-2}}$, consistent with zero, and a linear term likewise consistent with zero. 

We attribute this difference to the degeneracy between the orbital period and a systemic acceleration when the RVs span only $\sim$2 orbital cycles. Under that degeneracy an error in the period can accumulate as a slow drift in the predicted velocities that a low-order polynomial can absorb. The combined baseline of the RVs and astrometry time series here breaks this directly by anchoring the ephemeris 7.5 years before the RV epoch from \citet{barbato_gaps_2026}. This can be seen by the difference in the estimated period between our works (\autoref{tab:BinaryRVFit}) of $\sim$ 12.7 days or 2.2\% of an orbital cycle, which while negligible over a single cycle, over the temporal gap between the astrometry and RVs compounds to $\sim$ 10\% of an orbit (or roughly $40^{\circ}$ of orbital phase). 

\subsection{Astrometric and RV Jitter}\label{sec:astro_rvjitter}

We see the impact of the astrometric orbit on the residuals to the single star along-scan astrometry fit in \autoref{fig:Astrometry} and their RMS of 0.18 mas. Adding the astrometric orbit reduces this RMS to 0.100 mas, which compared to the median along-scan measurement error of 0.082 mas necessitates the need for the astrometric jitter term in our model. We constrain the log astrometric jitter to be $-3.00^{+0.08}_{-0.09}$ or $\sim$ 0.049 mas, but cannot determine the source of this jitter, whether it is centroiding uncertainties, calibration residuals or a manifestation of astrophysical jitter due to the star.  We note that a Lomb-Scargle periodogram \citep{lomb_least-squares_1976, scargle_studies_1982} on the along-scan residuals does not reveal any significant periodicities.

As noted by \citet{stefansson_gaia-4b_2025} as well, the RV jitters for HPF, NEID, FIES, (and now also HARPS-N) are higher than the individual instrument median RV uncertainties. The high RV jitter across the four instruments suggests that is not instrumental in origin, and could be astrophysical. This could be due to rotational modulation driven by spots coming in and out of view given the host star's relatively short photometric rotation period of $\sim$ 13.3 days \citep{stefansson_gaia-4b_2025}, or potentially additional companions in the system and should be explored in future works.


\begin{acknowledgments}
We thank the anonymous referee for a careful and constructive review that improved the clarity and quality of this manuscript.

SK acknowledges the use of Anthropic Claude AI Opus 5 to assist in (i) identifying bugs while adding an astrometric component to existing \texttt{exoplanet} inference scripts, particularly with $\omega$ and $\Omega$ sign degeneracies (ii) converting the posteriors to a formatted \latex table, (iii) coding up equations in latex, (iv) troubleshooting \texttt{matplotlib} for the astrometric result \autoref{fig:Astrometry}, and (v) documenting the code for \texttt{pyGaiafits}. SM acknowledges the use of Anthropic Claude AI Fable 5 for early development tests of code for fitting of Gaia astrometry and radial velocities, parts of which are incorporated into our pipeline. SK acknowledges responsibility for manually verifying the accuracy of all these outputs. 

This research has made use of the Exoplanet Follow-up Observation Program (ExoFOP; DOI: 10.26134/ExoFOP5) website, which is operated by the California Institute of Technology, under contract with the National Aeronautics and Space Administration under the Exoplanet Exploration Program. 

This work utilized (i) the NASA Exoplanet Archive, operated by Caltech under contract with NASA through the Exoplanet Exploration Program; (ii) the SIMBAD database, operated at CDS, Strasbourg, France; (iii) NASA’s Astrophysics Data System Bibliographic Services; and (iv) data from the Two Micron All Sky Survey (2MASS), a joint project of the University of Massachusetts and IPAC/Caltech, funded by NASA and the NSF.

This work has made use of data from the European Space Agency (ESA) mission
{\it Gaia} (\url{https://www.cosmos.esa.int/gaia}), processed by the {\it Gaia}
Data Processing and Analysis Consortium (DPAC,
\url{https://www.cosmos.esa.int/web/gaia/dpac/consortium}). Funding for the DPAC
has been provided by national institutions, in particular the institutions
participating in the {\it Gaia} Multilateral Agreement.

These results are based on observations obtained with the Habitable-zone Planet Finder Spectrograph on the HET. 

The Hobby-Eberly Telescope is a joint project of the University of Texas at Austin, the Pennsylvania State University, Ludwig-Maximilians-Universität München, and Georg-August Universität Gottingen. The HET is named in honor of its principal benefactors, William P. Hobby and Robert E. Eberly. The HET collaboration acknowledges the support and resources from the Texas Advanced Computing Center. We thank the Resident astronomers and Telescope Operators at the HET for the skillful execution of our observations with HPF. We would like to acknowledge that the HET is built on Indigenous land. Moreover, we would like to acknowledge and pay our respects to the Carrizo \& Comecrudo, Coahuiltecan, Caddo, Tonkawa, Comanche, Lipan Apache, Alabama-Coushatta, Kickapoo, Tigua Pueblo, and all the American Indian and Indigenous Peoples and communities who have been or have become a part of these lands and territories in Texas, on Turtle Island.

The Center for Exoplanets and Habitable Worlds is supported by Penn State and its Eberly College of Science.


M.Pi. acknowledges support from ASI-INAF agreement no. 2025-10-HH.0 “Partecipazione Italiana al Gaia DPAC - Supporto alle attività di responsabilità del team scientifico”. S.M., P.R., and G.S. are part of NASA's CHAMPs team, supported by NASA under grant No. 80NSSC23K1399 issued through the Interdisciplinary Consortia for Astrobiology Research (ICAR) program.

Resources supporting this work were provided by the (i) Pennsylvania State University's Institute for Computational and Data Sciences' (ICDS) Roar supercomputer, and (ii) the Resnick High Performance Computing Center, a facility supported by Resnick Sustainability Institute at the California Institute of Technology. This content is solely the responsibility of the authors and does not necessarily represent the views of the ICDS or Caltech. 

\end{acknowledgments}

\facilities{HET (HPF), WIYN (NEID), HARPS-North, FIES, Gaia}

\software{
\texttt{ArviZ} \citep{kumar_arviz_2019}, 
\texttt{astropy} \citep{robitaille_astropy:_2013, astropy_collaboration_astropy_2018, astropy_collaboration_astropy_2022},
\texttt{barycorrpy} \citep{kanodia_python_2018}, 
\texttt{exoplanet v0.4.4} \citep{foreman-mackey_exoplanet_2021, foreman-mackey_exoplanet-devexoplanet_2021},
\texttt{HxRGproc} \citep{ninan_habitable-zone_2018},
\texttt{ipython} \citep{perez_ipython:_2007},
\texttt{kima} python package \citep{faria_kima_2018}
\texttt{matplotlib} \citep{hunter_matplotlib:_2007},
\texttt{numpy} \citep{oliphant_numpy:_2006},
\texttt{pandas} \citep{mckinney-proc-scipy-2010},
\texttt{pyGaiafits} \citep[This work, ][]{kanodia_pygaiafits_2026}
\texttt{PyMC3} \citep{salvatier_probabilistic_2016},
\texttt{scipy} \citep{oliphant_python_2007, virtanen_scipy_2020},
\texttt{Theano} \citep{the_theano_development_team_theano_2016},
}

\bibliography{MyLibrary, references}{}
\bibliographystyle{aasjournalv7}



\end{document}